\documentclass[aps,prb,twocolumn,nobibnotes]{revtex4-1}  % for review and submission
\usepackage{dcolumn}
\usepackage{graphicx}
\usepackage[dvipsnames]{xcolor}
\usepackage{amssymb}
\usepackage{amsmath}
\usepackage{bm}
\usepackage[colorlinks=true,linktocpage=true,breaklinks=true]{hyperref}
\usepackage{multirow}

\begin{document}

\title{Orbital diffusion, polarization and swapping in centrosymmetric metals}
\author{Xiaobai Ning$^{1,2}$}
\author{A. Pezo$^{1,3}$}
\author{Kyoung-Whan Kim$^4$}
\author{Weisheng Zhao$^2$}
\author{Kyung-Jin Lee$^5$}
\author{Aur\'{e}lien Manchon$^1$}
\email{aurelien.manchon@univ-amu.fr}
\affiliation{$^1$Aix-Marseille Univ, CNRS, CINaM, Marseille, France; $^2$Fert Beijing Institute, School of Integrated Circuit Science and Engineering, National Key Laboratory of Spintronics, Beihang University, Beijing, China; $^3$Laboratoire Albert Fert, CNRS, Thales, Université Paris-Saclay, 91767, Palaiseau, France; $^4$Department of Physics, Yonsei University, Seoul, Korea; $^5$Department of Physics, Korea Advanced Institute of Science and Technology (KAIST), Daejeon 34141, Korea}

\begin{abstract}
We propose a general theory of charge, spin, and orbital diffusion based on Keldysh formalism. Our findings indicate that the diffusivity of orbital angular momentum in metals is much lower than that of spin or charge due to the strong orbital intermixing in crystals. Furthermore, our theory introduces the concept of “spin-orbit polarization” by which a pure orbital (spin) current induces a longitudinal spin (orbital) current, a process as efficient as spin polarization in ferromagnets. Finally, we find that orbital currents undergo momentum swapping, even in the absence of spin-orbit coupling. This theory establishes several key parameters for orbital transport of direct importance to experiments.
\end{abstract}

\maketitle
{\em Introduction - } The interconversion between charge and spin currents \cite{Nan2021} is one of the central mechanisms of spintronics and possibly its most instrumental. This mechanism is at the source of spin-orbit torque \cite{Manchon2019} and charge currents induced by spin pumping \cite{Saitoh2006}. At the core of these phenomena lies the spin-orbit interaction that couples the spin to the orbital angular momentum in high-Z materials (5$d$ metals, topological materials etc.). In recent years, it has been proposed that the interconversion between charge and orbital currents, via orbital Hall \cite{Bernevig2005b,Kontani2009,Go2018} and orbital Rashba effects \cite{Go2017,Yoda2018,Hamdi2023} for instance, might in fact be much more efficient than its spin counterpart because it arises from the orbital texture imposed by the crystal field rather than from spin-orbit coupling. Therefore, corresponding phenomena such as orbital torque \cite{Go2020,Ding2020,Lee2021b,Go2023} and orbital magnetoresistance \cite{Ding2022} have been proposed and experimentally reported. In these experiments, the scenario is based on a two-step process: orbital Hall or Rashba effect takes place in a light metal and the resulting orbital current is converted into a spin signal once in the adjacent ferromagnet. Consequently, it is expected that the supposedly large charge-to-orbital conversion taking place in the low-Z metal compensates for the relatively low spin-orbit coupling of the ferromagnet, which seems to be confirmed by the experiments \cite{Ding2020,Lee2021b,Ding2022,Sala2022,Hayashi2023,Fukunaga2023}. 

An important rationale behind the promotion of orbitronics is twofold. First, as mentioned above, since orbital transport is governed by the crystal field, orbital Hall and Rashba effects do not necessitate spin-orbit coupling and occur in low-Z metals (e.g., 3$d$ metals). Second, experiments suggest that orbital currents propagate over reasonably long distances ($\sim 10$ nm), which remains to be fully understood \cite{Lyalin2023,Choi2023,Hayashi2023,Fukunaga2023}. As a matter of fact, as observed by Ref. \onlinecite{Bernevig2005b}, the atomic orbital moment is never a good quantum number and therefore it remains unclear how it diffuses from one metal to another. Recent phenomenological models of orbital diffusion have been recently proposed \cite{Sala2022,Han2022} but lack quantitative predictability by overlooking microscopic details. In addition, several recent works have pointed out that the orbital moment arises not only from intra-atomic spherical harmonics ($p,\;d$) but also possesses substantial inter-atomic contribution \cite{Bhowal2021,Cysne2022,Pezo2022}. Understanding the way orbital currents and densities propagate in metals and accumulate at interfaces requires determining transport coefficients such as orbital conductivity or diffusivity, as well as the ability to interconvert spin currents into orbital currents via spin-orbit coupling. Indeed, when injecting an orbital density ${\bm l}=\langle \hat{\bf L}\rangle$ in metal, it diffuses and produces an orbital current ${\cal J}_l=-{\cal D}_l\partial_{\bm r}{\bm l}_l$, ${\cal D}_l$ being the orbital diffusion coefficient (typically a tensor). In the presence of spin-orbit coupling $\xi_{so}\hat{\bm\sigma}\cdot\hat{\bf L}$, this orbital current can convert into a spin current ${\cal J}_s$.

\begin{figure}[h!]
  \includegraphics[width=8cm]{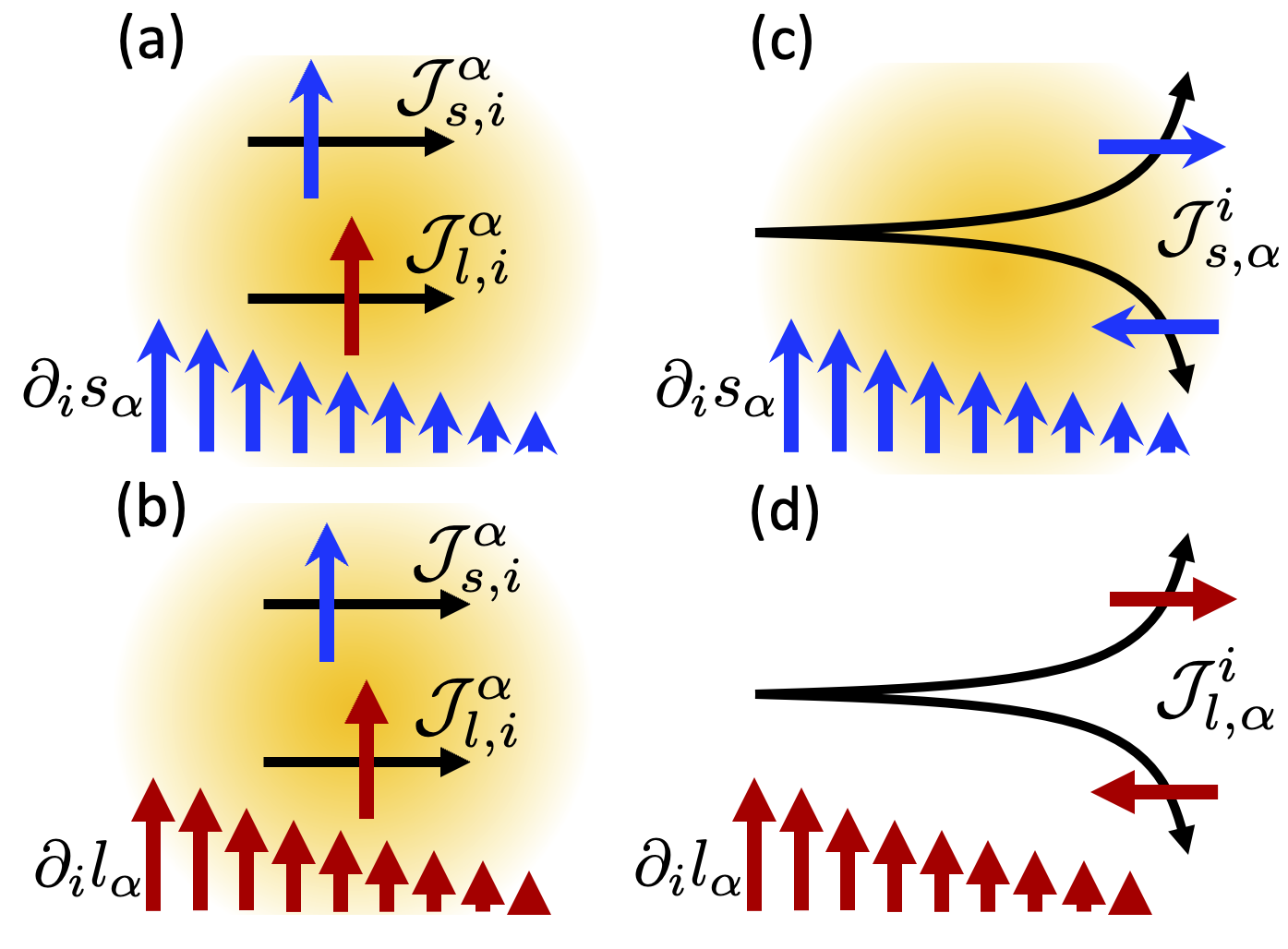}
  \caption{(Color online) Spin and orbit interconversion mechanims: (a) spin-to-orbit polarization and (b) orbit-to-spin polarization mediated by spin-orbit coupling. (c) Spin and (d) orbital swapping. The former requires spin-orbit coupling whereas the latter occurs even without it. The blue and red arrows represent the orbital and spin moment directions, and the black arrows represent the current propagation direction. The shaded region indicates the presence of spin-orbit coupling. \label{fig:sketch}}
\end{figure}

In this Letter, we derive a theory of spin and orbital diffusion in metals and uncover several mechanisms governing orbital torque and magnetoresistance phenomena, illustrated in Fig. \ref{fig:sketch}. First, we find that whereas charge and spin diffusion are of about the same order of magnitude, the orbital diffusion is much lower. This is due to the fact that the orbital moment is never a good quantum number in crystals (rotational invariance is broken). Second, we find that in the presence of spin-orbit coupling, an orbital current is systematically accompanied by a spin current that is collinear to it (and vice versa) [Fig. \ref{fig:sketch}(a,b)]. This "spin-orbit polarization" can be sizable, comparable to spin polarization in 3$d$ ferromagnets. Finally, the third class of effects uncovered by our theory is the "angular momentum swapping", i.e., the interchange between the propagation direction and angular momentum direction upon scattering [Fig. \ref{fig:sketch}(c,d)]. Whereas spin swapping was predicted by Lifshits and Dyakonov \cite{Lifshits2009} in the presence of spin-orbit coupling, orbital swapping arises naturally even without it. When turning on spin-orbit coupling, not only spin swapping emerges, but also spin-to-orbit and orbit-to-spin swapping.

{\em Theory - } The objective of the present theory is to determine the diffusive current induced by a gradient of particle density, ${\cal J}=-{\cal D}\partial_{\bf r}\rho$. In this expression, ${\cal J}$ can be the charge current $J_c$, or the spin (orbital) current ${\cal J}_{s(l)}$, whereas $\rho$ can be the charge density $\rho_c$, or the spin (orbital) density ${\bm s}$ (${\bm l}$). In the language of non-equilibrium Green's function, the particle current density is obtained by computing the quantum statistical expectation value of the trace of the particle current operator $\hat{j}$ taken over the lesser Green's function $G^<$,
\begin{equation}\label{eq:def1}
{\cal J}=\int \frac{d^3{\bf k}}{(2\pi)^3}\int \frac{d\varepsilon}{2i\pi}{\rm Tr}\left[\hat{j} G^<\right].
\end{equation}
The philosophy of the present theory is to express the lesser Green's function to the first order in the density gradient $\partial_{\bf r}\rho$. We start from the Keldysh-Dyson equation \cite{Rammer1986}
\begin{eqnarray}\label{eq:kd}
G^<=G^R\otimes\Sigma^<\otimes G^A,
\end{eqnarray}
where $G^<=G^<({\bf r},{\bf r}';t,t')$ ($\Sigma^<$) is the lesser Green's function (self-energy), $G^{R(A)}=G^{R(A)}({\bf r},{\bf r}';t,t')$ is the retarded (advanced) Green's function and $\otimes$ is the convolution product on both time and space. In Eq. \eqref{eq:kd}, we omitted the explicit time and space dependence for simplicity. In the linear response regime, we first express $G^<$ to the first order in spatial gradients using the Wigner transform (see, e.g., Ref. \onlinecite{SuppMat}), i.e., we rewrite Eq. \eqref{eq:kd} in the frame of the center-of-mass, $({\bf r},{\bf r}';t,t')\rightarrow ({\bf r}-{\bf r}',{\bf r}_c;t-t',t_c)$, with $({\bf r}_c,t_c)=(({\bf r}+{\bf r}')/2,(t+t')/2)$, Fourier transform the small space and time coordinates $({\bf r}-{\bf r}',t-t')\rightarrow ({\bf k},\omega)$, and expand Keldysh-Dyson equation to the first order in space and time gradients $(\partial_{{\bf r}_c},\partial_{t_c})$. In the following, the subscript $_c$ is dropped for the sake of readability. Under the Wigner transform, the convolution product becomes \cite{SuppMat}
\begin{eqnarray}\label{eq:conv}
A\otimes B=AB+\frac{i}{2}(\partial_{\bf r}A\partial_{\bf k}B-\partial_{\bf k}A\partial_{\bf r}B),
\end{eqnarray}
and finally, the part of the lesser Green's function that is linear in spatial gradient reads
\begin{eqnarray}\label{eq:Gless1}
\delta G^<=\frac{i}{2}\left(G_0^R\partial_{\bf r}\Sigma^<\partial_{\bf k}G_0^A-\partial_{\bf k}G_0^R\partial_{\bf r}\Sigma^<G_0^A\right).
\end{eqnarray}
Here $G_0^{R(A)}=(\hbar\omega-{\cal H}_0\pm i\Gamma)$ is the unperturbed retarded (advanced) Green's function and ${\cal H}_0$ is the crystal Hamiltonian. 

Since we are interested in the diffusion coefficients that connect angular momentum densities (odd under time-reversal $\mathcal{T}$) with angular momentum currents (even under $\mathcal{T}$), the diffusion coefficients in {\em nonmagnetic} materials are themselves odd under $\mathcal{T}$. The same is true for the charge diffusivity that connects the charge density (even under $\mathcal{T}$) with the charge current (odd under $\mathcal{T}$). As a result, the charge, spin, and orbital diffusion coefficients must be dissipative, proportional to the scattering time. In the language of quantum transport, these phenomena are driven by Fermi surface electrons akin to the charge conductivity. This is in stark contrast with the spin and orbital Hall diffusivities, which connect charge densities (even under $\mathcal{T}$) with spin and orbital currents (even under $\mathcal{T}$): they are even under $\mathcal{T}$, independent on scattering in the limit of weak disorder, and associated with the Berry curvature \cite{Sinova2004,Bonbien2020}. Since we focus on angular momentum diffusion and spin-orbit interconversion, Eq. \eqref{eq:Gless1} is limited to transport at the Fermi level and disregards Fermi sea contributions. The present analysis applies to nonmagnetic materials and must be revised in the case of magnetic systems \cite{Park2022b} as new terms are allowed.\par

Considering point-like impurities, $H_{\rm imp}=\sum_iV_0\delta({\bf r}-{\bf R}_i)$, the lesser self-energy reads
\begin{eqnarray}\label{eq:self}
\Sigma^<&=&\frac{1}{V}\sum_{i,j}\int\frac{d^3{\bf k}}{(2\pi)^3}V_0 G_{\bf k}^<V_0 e^{i{\bf k}\cdot({\bf R}_i-{\bf R}_j)}=\frac{n_iV_0^2}{V}\langle G_{\bf k}^<\rangle.\nonumber\\
\end{eqnarray}
Here, $n_i$ is the impurity concentration and $\langle...\rangle=V\int\frac{d^3{\bf k}}{(2\pi)^3}$ stands for momentum integration over the Brillouin zone. Noting that $\partial_{\bf k}G_0^R=\hbar G_0^R{\bf v}G_0^R$, we obtain
\begin{eqnarray}\label{eq:Gless2}
\delta G^<_{\bf k}=i\hbar\frac{n_iV_0^2}{V}{\rm Re}\left[G_0^R\partial_{\bf r}\langle G_{\bf k}^<\rangle G_0^A{\bf v}G_0^A\right].
\end{eqnarray}
Inserting Eq. \eqref{eq:Gless2} into Eq. \eqref{eq:def1}, we obtain the general expression of nonequilibrium properties induced by spatial gradients. Now, as argued above, diffusive effects are associated with Fermi surface electrons, which suggests $(1/V)\partial_{\bf r}\langle G_{\bf k}^<\rangle=2i\pi\partial_{\bf r}\hat{\rho}\delta(\varepsilon-\varepsilon_F)$, $\hat{\rho}$ being the density matrix at Fermi level. As a result, the particle current density reads
\begin{eqnarray}\label{eq:current}
{\cal J}=-\hbar n_iV_0^2{\rm ReTr}_{\bf k}\left[\hat{j} {\rm Im}[G_0^R\partial_{\bf r}\hat{\rho} G_0^A{\bf v}G_0^A]\right]_{\varepsilon_F}.
\end{eqnarray}
For the sake of compactness, we defined ${\rm Tr}_{\bf k}=\int\frac{d^3{\bf k}}{(2\pi)^3}{\rm Tr}$. Equation \eqref{eq:current} is the central result of this work and can be used to compute the diffusive charge, spin, and orbital currents induced by density gradients. For instance, substituting $\hat{\rho}$ by the charge density $\rho_c=-e{\rm Tr}[\hat{\rho}]$, and the charge current operator $\hat{j}=-e\hat{\bf v}$, one obtains the charge diffusivity
\begin{eqnarray}\label{eq:diffc}
{\cal D}_{ij}=\hbar n_iV_0^2{\rm ReTr}_{\bf k}\left[\hat{v}_j {\rm Im}[G_0^RG_0^A{\hat v}_iG_0^A]\right].
\end{eqnarray}
The validity of Eq. \eqref{eq:diffc} is readily assessed by comparing ${\cal D}_{ij}$ with the conductivity $\sigma_{ij}$ obtained from Kubo's formula \cite{Bonbien2020}. In the relaxation time approximation, $n_iV_0^2=\Gamma/(\pi{\cal N}_F)$, where ${\cal N}_F$ is the density of states at Fermi level and $\Gamma$ is the disorder broadening. Using this relation, we confirm the Einstein relation ${\cal D}_{ij}=\sigma_{ij}/(e^2 {\cal N}_F)$ \cite{SuppMat}. In the rest of this work, we express the spin and orbital diffusivity in the units of a conductivity $(e^2\Gamma/(\pi n_iV_0^2)){\cal D}_{ij}$, i.e., in $\Omega^{-1}\cdot m^{-1}$ rather than in $m^2\cdot s^{-1}$. 

To obtain the spin and orbital diffusivities, the respective densities are defined ${\bm s}=(\hbar/2){\rm Tr}[\hat{\bm\sigma}\hat{\rho}]$ and ${\bm l}=\hbar{\rm Tr}[\hat{\bf L}\hat{\rho}]$, $\hat{\bm\sigma}$ and $\hat{\bf L}$ being the dimensionless spin and orbital operators. Therefore, substituting the current operator $\hat{j}$ by either the spin current operator $\hat{j}_{s,j}^\alpha=(\hbar/4)\{\hat{v}_j,\hat{\sigma}_\beta\}$ or the orbital current operator $\hat{j}_{l,j}^\beta=(\hbar/2)\{\hat{v}_j,\hat{L}_\beta\}$ in Eq. \eqref{eq:current}, and $\partial_i\hat{\rho}$ by $\hat{\sigma}_\alpha\partial_i s_\alpha$ or $\hat{L}_\alpha\partial_{i}l_\alpha$, we obtain the general relation

\begin{eqnarray}\label{eq:diffmat}
\left(\begin{matrix}
{\cal J}_{s,j}^{\beta}\\
{\cal J}_{l,j}^{\beta}
\end{matrix}\right)=-\left(\begin{matrix}
{\cal D}_{s_\alpha i}^{s_\beta j} & {\cal D}_{l_\alpha i}^{s_\beta j}\\
{\cal D}_{s_\alpha i}^{l_\beta j} &{\cal D}_{l_\alpha i}^{l_\beta j}
\end{matrix}\right)\left(\begin{matrix}
\partial_i s_\alpha\\
\partial_i l_\alpha
\end{matrix}\right)
\end{eqnarray}

with the diffusion coefficients

\begin{eqnarray}\label{eq:diffc21}
{\cal D}_{s_\alpha i}^{s_\beta j} =2 n_iV_0^2{\rm ReTr}_{\bf k}\left[\hat{j}_{s,j}^\beta {\rm Im}[G_0^R\hat{\sigma}_\alpha G_0^A{\hat v}_iG_0^A]\right],\\\label{eq:diffc22}
{\cal D}_{l_\alpha i}^{l_\beta j}= n_iV_0^2{\rm ReTr}_{\bf k}\left[\hat{j}_{l,j}^\beta {\rm Im}[G_0^R\hat{L}_\alpha G_0^A{\hat v}_iG_0^A]\right],\\\label{eq:diffc23}
{\cal D}_{s_\alpha i}^{l_\beta j}=2 n_iV_0^2{\rm ReTr}_{\bf k}\left[\hat{j}_{l,j}^\beta {\rm Im}[G_0^R\hat{\sigma}_\alpha G_0^A{\hat v}_iG_0^A]\right],\\\label{eq:diffc24}
{\cal D}_{l_\alpha i}^{s_\beta j}= n_iV_0^2{\rm ReTr}_{\bf k}\left[\hat{j}_{s,j}^\beta {\rm Im}[G_0^R\hat{L}_\alpha G_0^A{\hat v}_iG_0^A]\right].\end{eqnarray}

${\cal D}_{s_\alpha i}^{s_\beta j}$ represents a spin current ${\cal J}_{s,j}^{\beta}$ induced by the gradient of a spin density $\partial_i s_\alpha$, whereas ${\cal D}_{l_\alpha i}^{l_\beta j}$ represents an orbital current ${\cal J}_{l,j}^{\beta}$ induced by the gradient of an orbital density $\partial_i l_\alpha$. In addition, the diffusivities ${\cal D}_{s_\alpha i}^{l_\beta j}$ and ${\cal D}_{l_\alpha i}^{s_\beta j}$ represent the spin-to-orbital interconversion phenomena, i.e., spin-orbit polarization ($\alpha=\beta$) and spin-orbit swapping ($\alpha\neq\beta$). 

\begin{figure}[h!]
  \includegraphics[width=8.5cm]{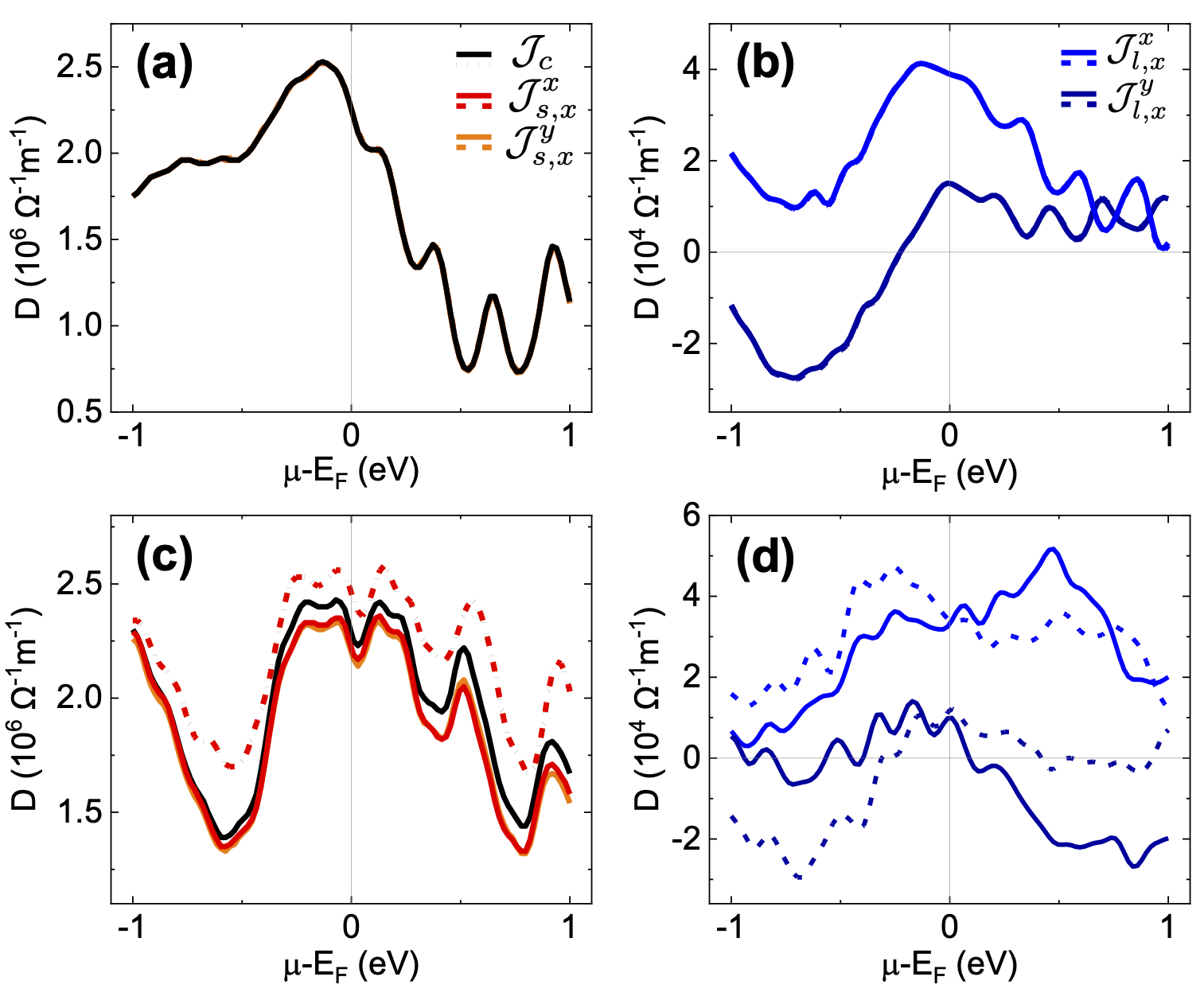}
  \caption{(Color online) (a, c) Charge (black), spin (red) and (b, d) orbital diffusivities as a function of the energy, without (dashed lines) and with (solid lines) spin-orbit coupling, for (a, b) V and (c, d) Ta. The spin and orbital diffusivities are computed for both $s_x,\;l_x$ (dark red, dark blue) and $s_y,\;l_y$ components (light red, light blue). The orbital diffusivities are strongly anisotropic and two orders of magnitude smaller than the spin and charge diffusivities.\label{fig:cond}}
  \end{figure}

{\em Spin and orbital diffusion -} To quantitatively estimate the magnitude of these effects in transition metals, we consider bcc crystalline structures with s, p, and d orbitals. An 18$\times$18 tight-binding Hamiltonian is obtained using the two-center approximation of Slater-Koster parameterization \cite{Slater1954}. Parameters up to the second-nearest neighbor are applied for V ($\xi_{\rm so}\approx 0$) and Ta ($\xi_{\rm so}\neq 0$) \cite{Papaconstantopoulos2015}, which belong to Group V in the periodic table and consequently share similar electronic properties.  Spin-orbit coupling is then added to the on-site energy \cite{Jaffe1987}, with the operator $\hat{\bf L}$ defined in Ref. \onlinecite{Dai2008b}. The resulting band structures are benchmarked against density functional theory \cite{SuppMat}. To compute the transport properties, we set the disorder broadening to $\Gamma=0.1$ eV and calculate the diffusion coefficients using a 50$\times$50$\times$50 $k$-grid. A toy model featuring a bcc crystal with $(p_x,p_y,p_z)$ orbitals is given in the Supplemental Materials \cite{SuppMat} and qualitatively confirms the conclusions drawn from these realistic calculations.

%{\color{blue}To quantitatively estimate the magnitude of these effects, we consider a bcc crystal with $(p_x,p_y,p_z)$ orbitals. The tight-binding Hamiltonian is obtained using Slater-Koster parameterization, with $V_\sigma=0.2$ eV and $V_\pi=0.05$ eV. Since the structure has cubic symmetry, we assume that the (charge, spin or orbital) gradient is along $x$.}

We first compute the charge, spin, and orbital diffusivities in Fig. \ref{fig:cond}. Since the current diffuses in the same direction as the density gradient, $i=j$, its angular momentum is necessarily aligned with that of the density, $\alpha=\beta$. In vanadium, with vanishingly small spin-orbit coupling, we find that the charge diffusivity ${\cal D}_{xx}$ and the spin diffusivities ${\cal D}_{s_x x}^{s_x x}$, ${\cal D}_{s_y x}^{s_y x}$ ($={\cal D}_{s_z x}^{s_z x}$) are all equal [Fig. \ref{fig:cond}(a)]. In tantalum [Fig. \ref{fig:cond}(b)], spin and charge diffusivities are slightly different due to the finite spin-orbit coupling, and ${\cal D}_{s_x x}^{s_x x}\approx{\cal D}_{s_y x}^{s_y x}(={\cal D}_{s_z x}^{s_z x})$. Interestingly, as reported on Fig. \ref{fig:cond}(b,d), the orbital diffusivities ${\cal D}_{l_x x}^{l_x x}\neq{\cal D}_{l_y x}^{l_y x}(={\cal D}_{l_z x}^{l_z x})$ are strongly anisotropic (dark and light blue lines) and, most importantly, are much smaller than the spin and charge diffusivities, which we attribute to the strong orbital mixing that naturally governs the band structure of our bcc crystal. In fact, the orbital angular momentum operator does not commute with the crystal Hamiltonian, $[\hat{\bf L},{\cal H}_0]\neq0$, resulting in its non-conservation: the electron's orbital moment is transferred to the lattice via the crystal field, exerting a mechanical torque. This non-conservation leads to the strong diffusion, and weak diffusion coefficient, of the orbital moment. In tantalum, spin-orbit coupling has a strong impact on the orbital diffusivity (dashed and solid lines). This can be understood qualitatively by the fact that spin-orbit interaction couples the highly conductive spin channel with the weakly conductive orbital channel, thereby (slightly) reducing the spin diffusivity while enhancing the orbital one, as shown in Fig. \ref{fig:cond}(c,d). Finally, let us point out that the orbital diffusivity changes sign across the band structure, in contrast to the charge and spin diffusivities. This sign change can be readily attributed to the oscillation of the orbital moment during the diffusion. This oscillation is associated with the dephasing between the orbital-quenched states whose superposition forms the orbital moment \cite{Han2022}. On the contrary, the spin degree of freedom is well-defined, only perturbed by spin-orbit coupling, and therefore its diffusivity follows that of the charge \cite{SuppMat}.
%$${\cal J}_{c}$$
%$${\cal J}_{s,x}^x$$
%$${\cal J}_{s,x}^y$$
%$${\cal J}_{l,x}^x$$
%$${\cal J}_{l,x}^y$$

The low orbital diffusivities reported here are consistent with the values reported by Sala et al. \cite{Sala2023}. Using Hanle effect, the authors estimate that the orbital diffusivity is about ${\cal D}_o\approx2.5\times10^{-6}\;m^2/s$ in Mn, corresponding to an orbital conductivity of $\sigma_o\approx5.5\times10^3\;\Omega^{-1}\cdot m^{-1}$. This experiment suggests that the orbital conductivity is two orders of magnitude smaller than the charge conductivity ($\sigma_c\approx6\times10^5\;\Omega^{-1}\cdot m^{-1}$ in Mn), which corroborates with our prediction.

{\em Spin-orbit polarization -} The next question we wish to address is how much orbital current can one obtain upon injecting a spin current in a heavy metal. This mechanism underlies the phenomena of orbital torque and orbital magnetoresistance\cite{Go2020,Ding2020,Lee2021b,Ding2022,Sala2022,Hayashi2023,Fukunaga2023} where a primary orbital current generated in a light metal is injected in a spin-orbit coupled material and converted into a spin current. To answer this question, we compute the so-called "spin-orbit polarization". Let us assume that a gradient of, say, spin density $\partial_i s_\alpha$ diffuses in the system. This gradient induces {\em both} spin and orbital currents, ${\cal J}_{s,i}^\alpha$ and ${\cal J}_{l,i}^\alpha$, producing a current of total angular momentum ${\cal J}_{t,i}^\alpha={\cal J}_{l,i}^\alpha+{\cal J}_{s,i}^\alpha$. To quantify the relative proportion of spin and orbital currents, we define the spin-to-orbit polarization ${\cal P}_{l,i}^\alpha=|{\cal D}_{s_\alpha i}^{l_\alpha i}|/(|{\cal D}_{s_\alpha i}^{s_\alpha i}|+|{\cal D}_{s_\alpha i}^{l_\alpha i}|)$, and similarly, the orbit-to-spin polarization, ${\cal P}_{s,i}^\alpha=|{\cal D}_{l_\alpha i}^{s_\alpha i}|/(|{\cal D}_{l_\alpha i}^{l_\alpha i}|+|{\cal D}_{l_\alpha i}^{s_\alpha i}|)$. Since the sign of the orbital current may vary across the band structure [e.g., Fig. \ref{fig:cond}(d)], only their absolute value enters the definition of the polarization. This way, the polarization assesses the efficiency of the conversion process itself.

\begin{figure}[h!]
  \includegraphics[width=8.5cm]{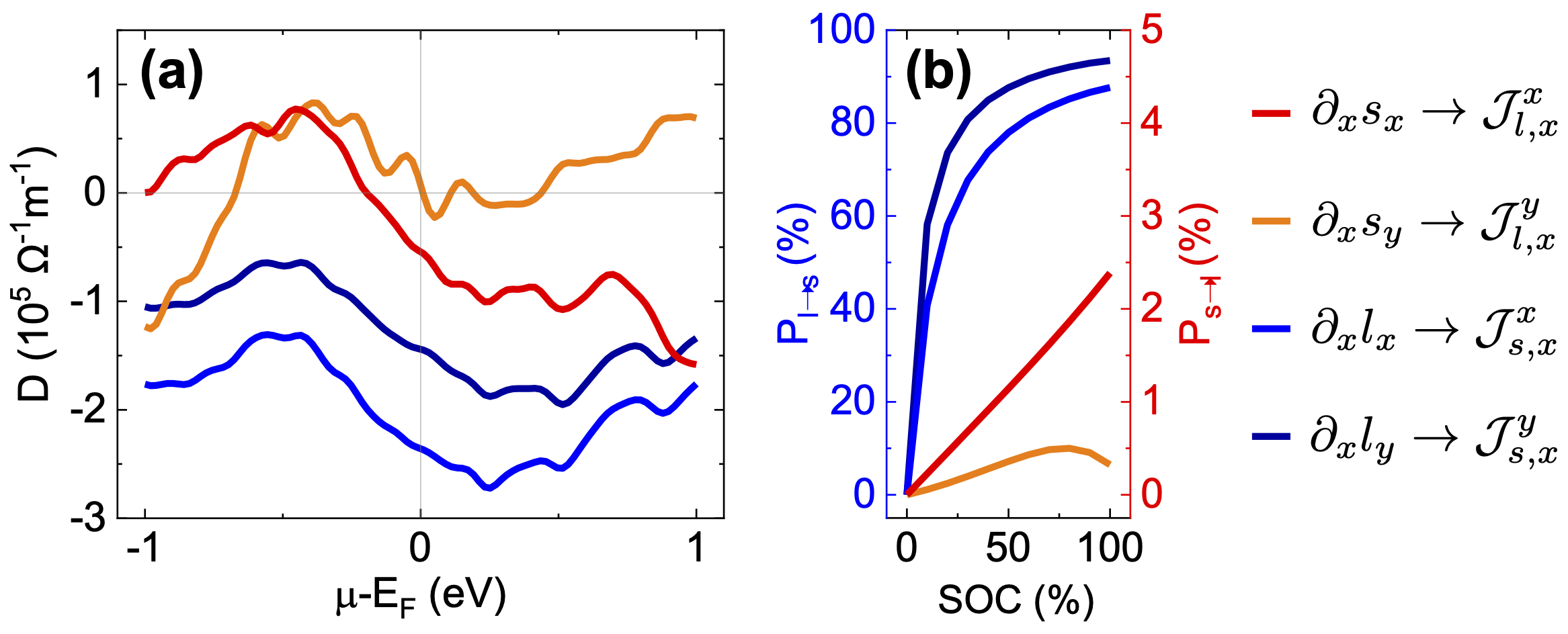}
  \caption{(Color online) (a) Spin-to-orbit (red) and orbit-to-spin (blue) diffusivities as a function of the energy in tantalum. (b) Corresponding spin-orbit polarizations as a function of the spin-orbit coupling. $100\%$ corresponds to the full spin-orbit coupling strength of tantalum. \label{fig:socpol}}
\end{figure}

The spin-to-orbit ($s_\alpha\rightarrow l_\alpha$) and orbit-to-spin ($l_\alpha\rightarrow s_\alpha$) longitudinal diffusivities in tantalum as well as the corresponding polarization are given in Fig. \ref{fig:socpol}(a) and (b), respectively. In this material, spin-to-orbit (red line) and orbit-to-spin diffusion coefficients (blue lines) are anisotropic due to the presence of spin-orbit coupling and of similar magnitude. The orbital diffusivity being much smaller than the spin diffusivity (see Fig. \ref{fig:cond}), the orbit-to-spin polarization is generally larger than the spin-to-orbit polarization (see further discussion in Ref. \cite{SuppMat}). As shown in Fig. \ref{fig:socpol}(b), the polarization increases steadily with spin-orbit coupling, as expected, and saturates at large spin-orbit coupling strength. It is worth noting that the orbit-to-spin polarization is comparable to the spin polarization found in conventional 3$d$ metal compounds (typically 50-70\%). 
%This observation is consistent with Ref. \onlinecite{Kontani2009} that suggests an orbit-to-spin polarization of about 50\% in Pt and Pd for Hall currents. 
The sizable spin-orbit polarization given in Fig. \ref{fig:socpol}(b) is a crucial ingredient for the orbital torque and magnetoresistance.

\begin{figure}[h!]
  \includegraphics[width=8.5cm]{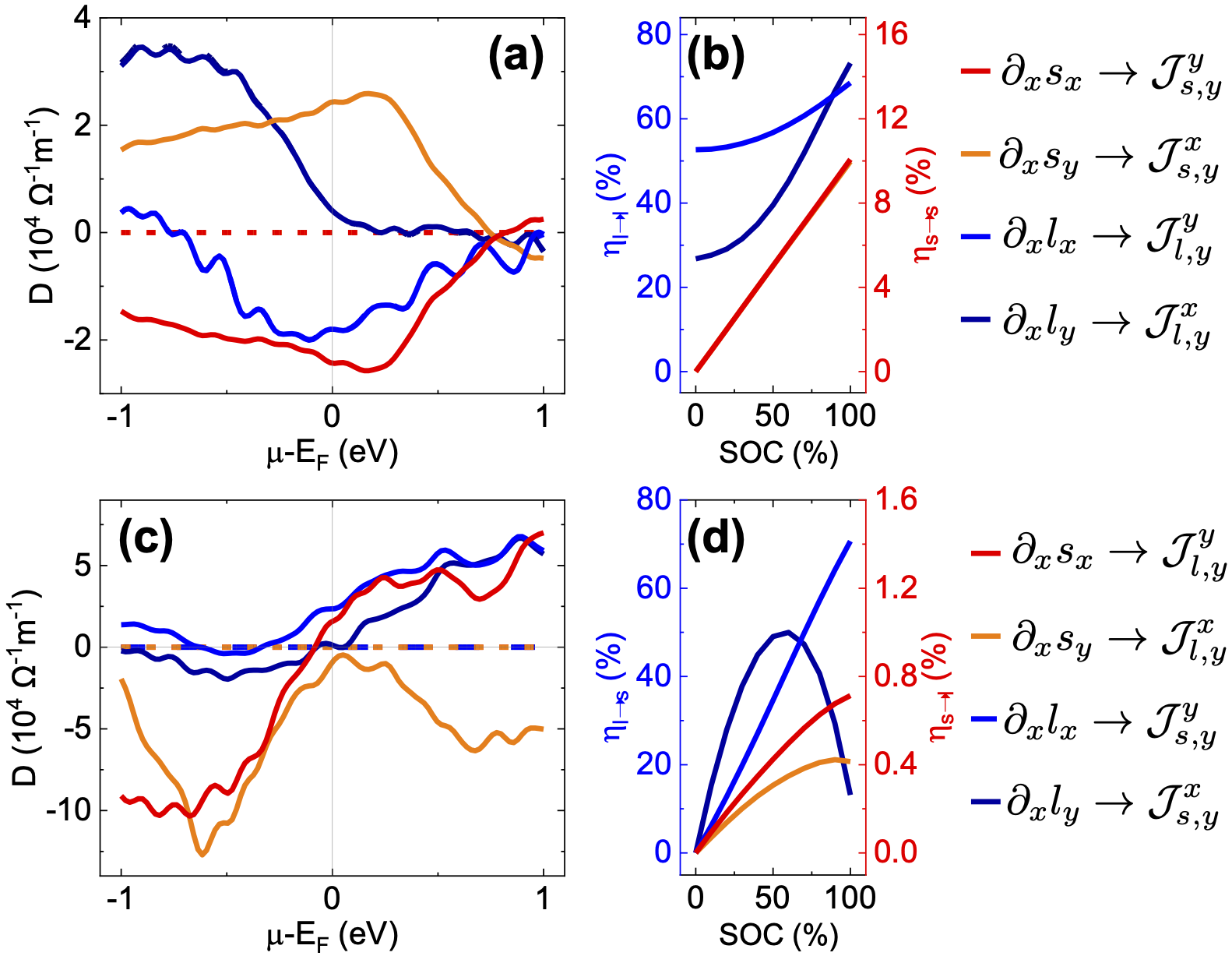}
  \caption{(a) Spin (red), orbital (blue), and (c) spin-to-orbit (red) and orbit-to-spin (blue) swapping diffusivities as a function of the energy in tantalum, without
(dashed lines) and with (solid lines) spin-orbit coupling. Corresponding (b) spin, orbit, and (d) spin-orbit swapping efficiencies as a function
of the spin-orbit coupling. 100\% corresponds to the full spin-orbit coupling strength of tantalum. \label{fig:swapping}}
\end{figure}

{\em Spin, orbital and spin-orbit swapping -} We finally consider the last class of effects, the spin and orbital swapping. For these effects, the directions of injection and collection are perpendicular to each other, as well as the direction of the incoming and outgoing (spin/orbit) polarization [see Fig. \ref{fig:sketch}(c,d)]. The orbital diffusivity tensor has the following form
\begin{eqnarray}\label{eq:tensorOrb}
\left(\begin{matrix}
{\cal J}_{l,x}^{x}\\
{\cal J}_{l,x}^{y}\\
{\cal J}_{l,y}^{x}\\
{\cal J}_{l,y}^{y}
\end{matrix}\right)=-\left(\begin{matrix}
{\cal D}_{l_x x}^{l_x x} & 0 & 0 & {\cal D}_{l_y y}^{l_x x}\\
0 & {\cal D}_{l_y x}^{l_y x} & {\cal D}_{l_x y}^{l_y x} & 0\\
0 & {\cal D}_{l_y x}^{l_x y} & {\cal D}_{l_x y}^{l_x y} & 0\\
{\cal D}_{l_x x}^{l_y y} & 0 & 0 & {\cal D}_{l_y y}^{l_y y}
\end{matrix}\right)\left(\begin{matrix}
\partial_x l_x\\
\partial_x l_y\\
\partial_y l_x\\
\partial_y l_y
\end{matrix}\right),
\end{eqnarray}
and Onsager reciprocity imposes that ${\cal D}_{l_x y}^{l_y x}={\cal D}_{l_y x}^{l_x y}$ and ${\cal D}_{l_y y}^{l_x x}={\cal D}_{l_x x}^{l_y y}$. Importantly, the orbital swapping does not necessitate spin-orbit coupling as it is solely governed by the orbital overlap (and hence the crystal field symmetry) of the crystal. These coefficients are reported in Fig. \ref{fig:swapping}(a) for tantalum, without (dashed lines) and with (solid lines) spin-orbit coupling. Without spin-orbit coupling, only the orbital swapping is allowed (blue lines). Turning on the spin-orbit coupling triggers spin swapping \cite{Lifshits2009} (red lines), whose diffusivity tensor has the same form as in Eq. \eqref{eq:tensorOrb}. Figure \ref{fig:swapping}(b) displays the spin (red) and orbital (blue) swapping efficiencies defined as $\eta_{s\rightarrow s}=|{\cal D}_{s_xx}^{s_y y}/{\cal D}_{s_xx}^{s_x x}|$ and $\eta_{l\rightarrow l}=|{\cal D}_{l_xx}^{l_y y}/{\cal D}_{l_xx}^{l_x x}|$, as a function of spin-orbit coupling, showing that orbital swapping is generally larger than spin swapping, which is reasonable given the minor role of spin-orbit coupling in the former.

In addition, spin-orbit coupling also enables the transfer between spin and orbital angular momenta that results in spin-to-orbit (red) and orbit-to-spin (blue) swapping, displayed in Fig. \ref{fig:swapping}(c). The diffusivity tensor has the form

\begin{eqnarray}\label{eq:diffmat}
\left(\begin{matrix}
{\cal J}_{s,x}^{x}\\
{\cal J}_{s,x}^{y}\\
{\cal J}_{s,y}^{x}\\
{\cal J}_{s,y}^{y}
\end{matrix}\right)=-\left(\begin{matrix}
{\cal D}_{l_x x}^{s_x x} & 0 & 0 & {\cal D}_{l_y y}^{s_x x}\\
0 & {\cal D}_{l_y x}^{s_y x} & {\cal D}_{l_x y}^{s_y x} & 0\\
0 & {\cal D}_{l_y x}^{s_x y} & {\cal D}_{l_x y}^{s_x y} & 0\\
{\cal D}_{l_x x}^{s_y y} & 0 & 0 & {\cal D}_{l_y y}^{s_y y}
\end{matrix}\right)\left(\begin{matrix}
\partial_x l_x\\
\partial_x l_y\\
\partial_y l_x\\
\partial_y l_y
\end{matrix}\right),
\end{eqnarray}
and Onsager reciprocity imposes ${\cal D}_{l_x y}^{s_y x}={\cal D}_{l_y x}^{s_x y}$ and ${\cal D}_{l_y y}^{s_x x}={\cal D}_{l_x x}^{s_y y}$. Swapping efficiencies, $\eta_{l\rightarrow s}$ and $\eta_{s\rightarrow l}$, are defined similarly as above. From Fig. \ref{fig:swapping}(c), it appears that although the spin-to-orbit swapping diffusivity (red) is larger than the orbit-to-spin swapping diffusivity (blue), the total efficiency of the orbital-to-spin swapping efficiency remains much larger due to the small orbital diffusivity [Fig. \ref{fig:swapping}(d)]. In the context of spin-orbit torque \cite{Manchon2019}, spin swapping, being of bulk \cite{Lifshits2009,Saidaoui2016} or interfacial origin \cite{Baek2018}, is responsible for additional torque components in magnetic multilayers, depending on the transport regime (diffusive versus Knudsen regime). The large orbital swapping efficiencies reported here suggest that in systems displaying orbital torque, deviations from the conventional damping-like torque can be expected \cite{Saidaoui2016}.

{\em Conclusion - } Advancing research in orbitronics requires a proper description of spin and orbital diffusion in metals. As stated previously, whereas the vast majority of theoretical studies to date focus on orbital and spin currents generated by electric currents, our theory allows us to compute the orbital and spin currents induced by diffusive gradients of angular momenta. It reveals that although orbital currents do not experience "orbital-flip" {\em per se}, their diffusivity in metals is much weaker than that of spin currents, in qualitative agreement with Ref. \cite{Sala2023}. This theory sets a milestone in the theoretical description of the orbital relaxation reported in low-Z metals\cite{Lee2021b,Hayashi2023,Fukunaga2023,Lyalin2023,Choi2023}. In diffusive transport, the (spin or orbital) diffusion length is related to the product between the (spin or orbital) diffusivity ${\cal D}$ and the (spin or orbital) relaxation time $\tau_r$, $\lambda\propto \sqrt{{\cal D}\tau_r}$. In a recent theory, Sohn et al. \cite{Sohn2024} suggests that the orbital relaxation time is driven by a D'yakonov-Perel mechanism, which constitutes an appealing direction for a comprehensive orbital transport theory. An important outcome of our theory is the calculation of the spin-to-orbit and orbit-to-spin polarizations, i.e., the ability of a spin (orbital) current to generate a longitudinal orbital (spin) current. We find that the orbital-spin polarization is very efficient, potentially as efficient as conventional spin polarization in 3$d$ magnets. Finally, we show that orbital currents are subject to angular moment swapping even in the absence of spin-orbit coupling and can be as large as spin swapping, and orbit-to-spin swapping is much more efficient than spin-to-orbit swapping due to the weak orbital diffusion coefficient.
%
%A comprehensive theory of orbital transport requires a microscopic modeling of orbital relaxation mechanisms, which is out of the scope of the present work \cite{Sohn2024}.} Notice that in ferromagnets, spin dephasing severely limits the spin propagation, such that orbital diffusion naturally dominates \cite{Lee2021b,Go2023}. This effect is absent in nonmagnetic metals.

%We point out that the Green function theory proposed in this Letter is well adapted to multiband systems and in particular to realistic heterostructures computed from first principles. Indeed, systematic investigation of orbital Hall conductivity and orbital Edelstein effects in transition metals have been recently performed \cite{Jo2018,Salemi2021,Salemi2022} and extending the present work to realistic materials of interest to experiments could open appealing perspectives for the design of orbital devices. 

\begin{acknowledgments}
X.N. appreciates Yuhao Jiang's advice and assistance during the code development phase. X.N. was supported by the China Scholarship Council Program. W.Z was supported by the National Key Research and Development Program of China (Grants No. 2022YFB4400200), the National Natural Science Foundation of China (Grant No. 92164206, No. 52121001, and No. 52261145694), and the New Cornerstone Science Foundation through the XPLORER PRIZE. A.P. was supported by the ANR ORION project, grant ANR-20-CE30-0022-01 of the French Agence Nationale de la Recherche, A. M. was supported by France 2030 government investment plan managed by the French National Research Agency under grant reference PEPR SPIN – [SPINTHEORY] ANR-22-EXSP-0009, and by the EIC Pathfinder OPEN grant 101129641 “OBELIX”. K.-W. K. was supported by the KIST Institutional Program and the National Research Foundation (NRF) of Korea (RS-2024-00334933, RS-2024-00410027), and K.-J. L. was supported by the NRF (NRF-2022M3I7A2079267).
\end{acknowledgments}

\bibliography{Biblio2023}

\end{document}

\title{Supplemental Materials: Orbital diffusion, polarization and swapping in centrosymmetric metals}

\author{Xiaobai Ning$^{1,2}$}
\author{A. Pezo$^{1,3}$}
\author{Kyoung-Whan Kim$^4$}
\author{Weisheng Zhao$^2$}
\author{Kyung-Jin Lee$^5$}
\author{Aur\'{e}lien Manchon$^1$}
\email{aurelien.manchon@univ-amu.fr}
\affiliation{$^1$Aix-Marseille Univ, CNRS, CINaM, Marseille, France; $^2$Fert Beijing Institute, School of Integrated Circuit Science and Engineering, National Key Laboratory of Spintronics, Beihang University, Beijing, China; $^3$Laboratoire Albert Fert, CNRS, Thales, Université Paris-Saclay, 91767, Palaiseau, France; $^4$Center for Spintronics, Korea Institute of Science and Technology, Seoul 02792, Korea; $^5$Department of Physics, Korea Advanced Institute of Science and Technology (KAIST), Daejeon 34141, Korea}

\maketitle
\section{Keldysh theory for diffusive currents}
As explained in the main text, our objective is to express generalized current densities induced by generalized chemical potential gradients, in a spirit akin to Onsager's. To do so, we start from Keldysh-Dyson equation \cite{Rammer1986,Mahan2000},
\begin{eqnarray}\label{eq:kd}
G^<({\bf r},{\bf r}';t,t')=\int G^R({\bf r},{\bf r}_1;t,t_1)\Sigma^<({\bf r}_1,{\bf r}_2;t_1,t_2) G^A({\bf r}_2,{\bf r}';t_2,t')d{\bf r}_1d{\bf r}_2dt_1dt_2,
\end{eqnarray}
where $G^<$ ($\Sigma^<$) is the lesser Green's function (self-energy), $G^{R(A)}$ is the retarded (advanced) Green's function. To solve this equation, we use Wigner transform that rewrites the Green functions in the frame of the center of mass, $({\bf r},{\bf r}';t,t')\rightarrow ({\bf r}-{\bf r}',{\bf r}_c;t-t',t_c)$, with $({\bf r}_c,t_c)=(({\bf r}+{\bf r}')/2,(t+t')/2)$. Once such a variable change has been achieved, one Fourier transform the small space and time coordinates $({\bf r}-{\bf r}',t-t')\rightarrow ({\bf k},\omega)$, and expand the Green functions to the lowest orders in space and time gradients $(\partial_{{\bf r}_c},\partial_{t_c})$. This method enables to express the convolution product as
\begin{eqnarray}\label{eq:conv}
A\otimes B=AB+\frac{i}{2}(\partial_{\bf r}A\partial_{\bf k}B-\partial_{\bf k}A\partial_{\bf r}B),
\end{eqnarray}

This procedure is exposed in detail in Refs. \cite{Rammer1986,Mahan2000,Wang2014} and summarized below.

\subsection{Wigner expansion}

Since the Wigner transform is a powerful but cumbersome procedure, we work on the 4-dimensional vector $x=({\bf r},t)$. Let us consider the following convolution product,
\begin{eqnarray}\label{eq:wign1}
C(x_1,x_2)=\int A(x_1,x')B(x',x_2)dx'.
\end{eqnarray}
We now rewrite the coordinates, $x_1=X+\frac{x}{2}$, $x_2=X-\frac{x}{2}$, where $X$ represents the position and time of the center of mass of the wave packet and $x$ represents the deviation from the center of mass. Then, by noting $\bar{C}(X,x)=C(X+\frac{x}{2},X-\frac{x}{2})$, Eq. \ref{eq:wign1} can be rewritten
\begin{eqnarray}\label{eq:wign1}
\bar{C}(X,x)&=&\int A\left(X+\frac{x}{2},x'\right)B\left(x',X-\frac{x}{2}\right)dx',\\
\bar{C}(X,x)&=&\int \bar{A}\left(\frac{X+\frac{x}{2}+x'}{2},X+\frac{x}{2}-x'\right)\bar{B}\left(\frac{X-\frac{x}{2}+x'}{2},x'-X+\frac{x}{2}\right)dx'.
\end{eqnarray}
Now, let's perform the change of variable $x'\rightarrow X-\frac{x}{2}+x'$, and we get
\begin{eqnarray}\label{eq:wign2}
\bar{C}(X,x)&=&\int \bar{A}\left(X+\frac{x'}{2},x-x'\right)\bar{B}\left(X-\frac{x}{2}+\frac{x'}{2},x'\right)dx'.
\end{eqnarray}
We now apply a Fourier transform on the small variables, $x$, $x'$ and $x-x'$, in order to only keep the coordinate of the center of mass, $X$, explicit. We then get
\begin{eqnarray}\label{eq:wign3}
C(X,p)&=&\int dx e^{-ipx} \bar{C}(X,x),\\\label{eq:wign3b}
C(X,p)&=&\int dxe^{-ipx}\int dx' \int \frac{dp'}{(2\pi)^4}\int \frac{dp''}{(2\pi)^4} e^{ip'(x-x')}A\left(X+\frac{x'}{2},p'\right)e^{ip''x'}B\left(X-\frac{x}{2}+\frac{x'}{2},p''\right)
\end{eqnarray}
Let us now perform a Taylor expansion on $A$ and $B$ with respect to $x'$ and $x'-x$. In principle, the procedure can be applied to all orders, but for the purpose of the present work, we only apply it to the first order. This expansion gives
\begin{eqnarray}\label{eq:wign4}
A\left(X+\frac{x'}{2},p'\right)B\left(X-\frac{x}{2}+\frac{x'}{2},p''\right)\approx \left(A(X,p')+\frac{x'}{2}\frac{\partial A}{\partial X}(X,p')\right)\left(B(X,p'')+\frac{x'-x}{2}\frac{\partial B}{\partial X}(X,p'')\right).
\end{eqnarray}
We therefore obtain the expression of term $C(X,p)$ to the first order in gradient by introducing Eq. \ref{eq:wign4} into Eq. \ref{eq:wign3b} and performing the integration. Finally, we obtain
\begin{eqnarray}\label{eq:wign5}
C(X,p)=A(X,p)B(X,p)-\frac{i}{2}\frac{\partial}{\partial p}A(X,p)\frac{\partial}{\partial X}B(X,p)+\frac{i}{2}\frac{\partial}{\partial X}A(X,p)\frac{\partial}{\partial p}B(X,p) 
\end{eqnarray}
In our system, conjugate variables are $x=({\bf r},t)$ and $p=(E,t)$ and therefore, $\frac{\partial}{\partial X}=({\bm\nabla}_{\bf r},-\partial_t)$ and
$\frac{\partial}{\partial p}=({\bm\nabla}_{\bf p},-\partial_E)$. In other words, the convolution product in the main text can be written,
\begin{eqnarray}\label{eq:convmain}
A\otimes B=AB+\frac{i}{2}(\partial_{\bf r}A\partial_{\bf k}B-\partial_{\bf k}A\partial_{\bf r}B),
\end{eqnarray}

\subsection{Green function at the first order in gradient}

Let us now apply the Wigner expansion, Eq. \eqref{eq:convmain}, to the Keldysh-Dyson equation \cite{Rammer1986}, Eq. (3) in the main text,
\begin{eqnarray}\label{eq:kd0}
G^<=G^R\otimes\Sigma^<\otimes G^A,\\
(\hbar\omega-{\cal H}_0-\Sigma^R)\otimes G^R=1,\\
G^A\otimes(\hbar\omega-{\cal H}_0-\Sigma^A)\otimes=1.
\end{eqnarray}
$\Sigma^{R,A,<}$ are the self-energies defined
\begin{eqnarray}\label{eq:self}
\Sigma^{R,A,<}&=&\frac{1}{V}\sum_{i,j}\int\frac{d^3{\bf k}}{(2\pi)^3}V_0 G_{\bf k}^{R,A,<}V_0 e^{i{\bf k}\cdot({\bf R}_i-{\bf R}_j)}=\frac{n_iV_0^2}{V}\langle G_{\bf k}^{R,A,<}\rangle.
\end{eqnarray}
Here, $n_i$ is the impurity concentration and $\langle...\rangle=V\int\frac{d^3{\bf k}}{(2\pi)^3}$ stands for momentum integration over the Brillouin zone. Notice that in the case of localized, delta-like impurity potential, these self-energies are momentum-independent (see the discussion in the main text). This expression involves two Moyal products, so we work on the right one first,
\begin{eqnarray}\label{eq:kd11}
G^<&=&G^R\otimes(\Sigma^<\otimes G^A),\\\label{eq:kd12}
G^<&=&G^R\otimes\left(\Sigma^<G^A+\frac{i}{2}(\partial_{\bf r}\Sigma^<\partial_{\bf k}G^A-\partial_{\bf k}\Sigma^<\partial_{\bf r}G^A)\right).
\end{eqnarray}
Since $\Sigma^<$ is momentum-independent, $\partial_{\bf k}\Sigma^<=0$ and we obtain,
\begin{eqnarray}\label{eq:kd21}
G^<&=&G^R\Sigma^<G^A+\frac{i}{2}G^R\partial_{\bf r}\Sigma^<\partial_{\bf k}G^A
+\frac{i}{2}\left[\partial_{\bf r}G^R\partial_{\bf k}\left(\Sigma^<G^A+\frac{i}{2}\partial_{\bf r}\Sigma^<\partial_{\bf k}G^A\right)
-\partial_{\bf k}G^R\partial_{\bf r}\left(\Sigma^<G^A+\frac{i}{2}\partial_{\bf r}\Sigma^<\partial_{\bf k}G^A\right)\right].\nonumber\\
\end{eqnarray}
Since we are interested in modeling the diffusive regime, i.e., the response of the system to a spatial gradient of particle density, we only retain the terms that are proportional to $\partial_{\bf r}\Sigma^<$. The other terms, proportional to $\Sigma^<$, involve interband transitions which are disregarded in the present theory. Therefore, we obtain the part of the less Green's function that is due to particle diffusion reads
\begin{eqnarray}\label{eq:kd21}
\delta G^<&=&\frac{i}{2}(G^R\partial_{\bf r}\Sigma^<\partial_{\bf k}G^A-\partial_{\bf k}G^R\partial_{\bf r}\Sigma^<G^A).
\end{eqnarray}
Since $\Sigma^<$ is already proportional to disorder, it is sufficient to take the unperturbed retarded and advanced Green's functions, $G^{A,R}\rightarrow G^{A,R}_0$. We finally arrive at Eq. (4) in the main text.

\section{Relationship between charge diffusivity and conductivity}

Let us demonstrate the validity of Einstein's relation between the charge diffusivity, Eq. (8), and the charge conductivity (see, e.g., Ref. \onlinecite{Bonbien2020}). We start from Eq. (8),
\begin{eqnarray}\label{eq:diffc}
{\cal D}_{ij}=\hbar n_iV_0^2{\rm ReTr}_{\bf k}\left[\hat{v}_j {\rm Im}[G_0^RG_0^A{\hat v}_iG_0^A]\right],
\end{eqnarray}
and recognize that for $\Gamma\rightarrow0$, $G_0^RG_0^A\approx -(1/i\Gamma)G_0^R=(1/i\Gamma)G_0^A$. First write the imaginary part ${\rm Im[...]}$ explicitly and then replacing $G_0^RG_0^A$ in Eq. \eqref{eq:diffc} leads to
\begin{eqnarray}\label{eq:diffc}
{\cal D}_{ij}&=&\hbar n_iV_0^2{\rm ReTr}_{\bf k}\left[(-i)\hat{v}_j[G_0^RG_0^A{\hat v}_iG_0^A-G_0^R{\hat v}_iG_0^RG_0^A]\right],\\
{\cal D}_{ij}&=&\frac{\hbar}{2} \frac{n_iV_0^2}{\Gamma}{\rm ReTr}_{\bf k}\left[\hat{v}_j[G_0^R{\hat v}_iG_0^A-G_0^R{\hat v}_iG_0^R-G_0^A{\hat v}_iG_0^A+G_0^R{\hat v}_iG_0^A]\right].\\
{\cal D}_{ij}&=&\frac{\hbar}{4} \frac{n_iV_0^2}{\Gamma}{\rm ReTr}_{\bf k}\left[\hat{v}_j[G_0^R{\hat v}_i(G_0^A-G_0^R)-(G_0^A-G_0^R){\hat v}_iG_0^A]\right].
\end{eqnarray}
Since ${\cal D}_{ij}$ is even under time-reversal symmetry (see main text), ${\cal D}_{ij}={\cal D}_{ji}$, which gives
\begin{eqnarray}
{\cal D}_{ij}&=&\frac{\hbar}{8} \frac{n_iV_0^2}{\Gamma}{\rm ReTr}_{\bf k}\left[\hat{v}_jG_0^R{\hat v}_i(G_0^A-G_0^R)-\hat{v}_j(G_0^A-G_0^R){\hat v}_iG_0^A+\hat{v}_iG_0^R{\hat v}_j(G_0^A-G_0^R)-\hat{v}_i(G_0^A-G_0^R){\hat v}_jG_0^A\right].
\end{eqnarray}
After some rearrangement, we finally obtain
\begin{eqnarray}
{\cal D}_{ij}&=&-\frac{\hbar}{4} \frac{n_iV_0^2}{\Gamma}{\rm ReTr}_{\bf k}\left[\hat{v}_j(G_0^A-G_0^R){\hat v}_i(G_0^A-G_0^R)\right].
\end{eqnarray}
Since $\Gamma/(n_iV_0^2)=\pi {\cal N}_F$ (see main text), we end up with
\begin{eqnarray}
{\cal D}_{ij}&=&-\frac{1}{e^2{\cal N}_F}\frac{\hbar e^2}{4} \frac{n_iV_0^2}{\Gamma}{\rm ReTr}_{\bf k}\left[\hat{v}_j(G_0^A-G_0^R){\hat v}_i(G_0^A-G_0^R)\right]=\frac{\sigma_{ij}}{e^2{\cal N}_F},
\end{eqnarray}
where the tensor $\sigma_{ij}$ is the time-reversal even component of the conductivity, as expressed in Ref. \onlinecite{Bonbien2020}. We recover the well-known Einstein formula.

\section{Computation of orbital diffusion coefficients in realistic metals}

In this section, we report the electronic band structure of the three selected transition metals, V, Nb and Ta, computed by DFT using QUANTUM ESPRESSO (solid black lines) ($k$-grid: 17$\times$17$\times$17) and using the tight-binding code (dashed red lines). The on-site energy at $\Gamma$-point is shifted when applying the tight-binding method, but it could also result from the selection of the PBE pseudopotential in DFT. The band structure near the Fermi level is consistent between the two methods, ensuring the validity of the diffusivity calculations.

\begin{figure}[h!]
  \includegraphics[width=15cm]{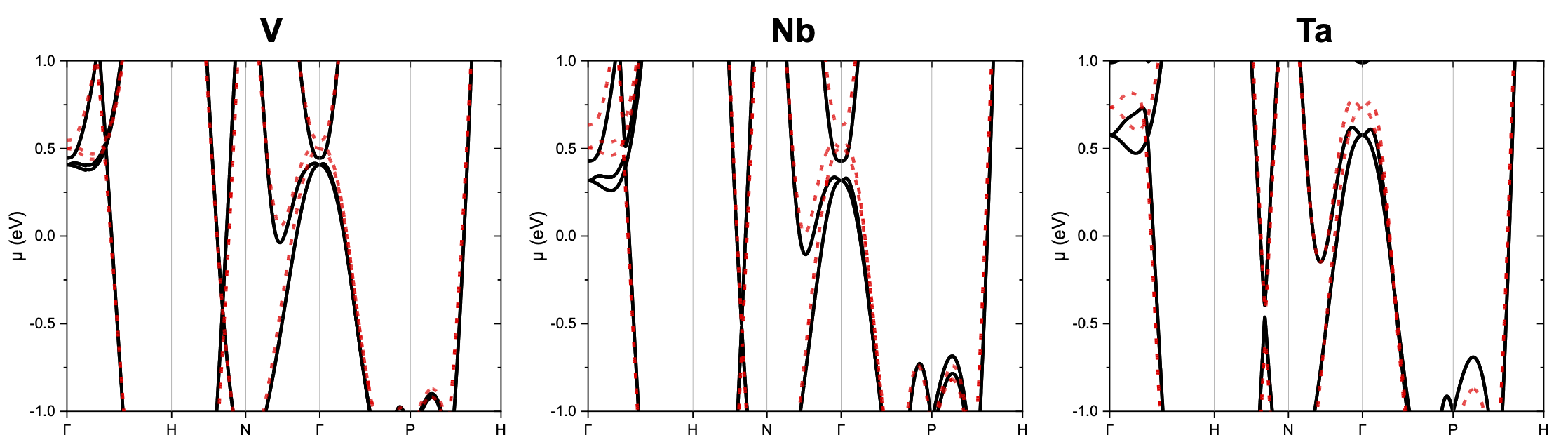}
  \caption{(Color online) Band structure of bcc vanadium (left), niobium (center) and tantalum (right) computed by DFT (solid black lines) and using tight-binding parameterization (dashed red lines).\label{fig:dft-tb}}
\end{figure}

\section{A minimal model for orbital diffusion}

\begin{figure}[h!]
  \includegraphics[width=8.5cm]{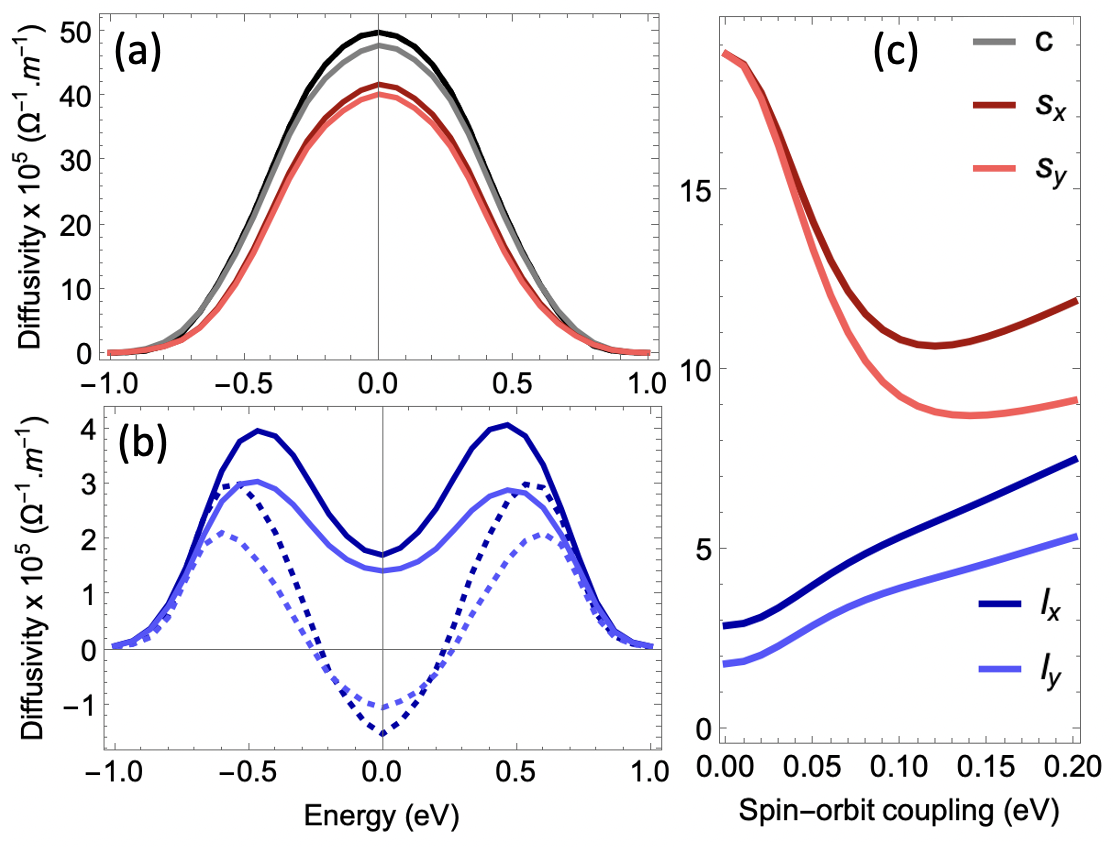}
  \caption{(Color online) (a) Charge (black and gray) and spin (red) diffusivities as a function of the energy. For $\xi_{\rm so}=0$, the spin and charge diffusivities fall into one single curve (black), whereas for $\xi_{\rm so}=0.05$, the charge diffusivity is reduced (gray) and the spin diffusivity becomes anisotropic (light and dark red lines). (b) Orbital diffusivities as a function of the energy for $\xi_{\rm so}=0$ (dashed) and $\xi_{\rm so}=0.05$ eV (solid). The orbital diffusivities are intrinsically anisotropic. (c) Dependence of the spin (red) and orbital (blue) diffusivities as a function of the spin-orbit coupling $\xi_{\rm so}$ at transport energy $\varepsilon=0.5$ eV.\label{fig:cond}}
\end{figure}

In this section we present the orbital transport properties of a toy model featuring a bcc crystal with $(p_x,p_y,p_z)$ orbitals with nearest-neighbor hopping. The tight-binding Hamiltonian is obtained using Slater-Koster parameterization, with $V_\sigma=0.2$ eV and $V_\pi=0.05$ eV. Since the structure has cubic symmetry, we assume that the (charge, spin, or orbital) gradient is along $x$. 

We first compute the charge, spin and orbital diffusivities in Fig. \ref{fig:cond}. Since the current diffuses in the same direction as the density gradient, $i=j$, its angular momentum is necessarily aligned on that of the density, $\alpha=\beta$. In the absence of spin-orbit coupling, the charge diffusivity ${\cal D}_{xx}$ and the spin diffusivities ${\cal D}_{s_x x}^{s_x x}$, ${\cal D}_{s_y x}^{s_y x}$ ($={\cal D}_{s_z x}^{s_z x}$) are all equal [black line in Fig. \ref{fig:cond}(a)]. Turning on the spin-orbit coupling ($\xi_{\rm so}=0.05$ eV) slightly reduces the charge diffusivity and breaks the symmetry between the spin diffusion coefficients, ${\cal D}_{s_x x}^{s_x x}\neq{\cal D}_{s_y x}^{s_y x}(={\cal D}_{s_z x}^{s_z x})$. Interestingly, as reported on Fig. \ref{fig:cond}(b), the orbital diffusivities ${\cal D}_{l_x x}^{l_x x}$ and ${\cal D}_{l_y x}^{l_y x}$ ($={\cal D}_{l_z x}^{l_z x}$) are in fact very small (dashed blue lines), which we attribute to the strong orbital mixing that naturally governs the band structure of our bcc crystal, as discussed in the main text. Turning on the spin-orbit coupling (solid lines) again breaks the symmetry between the diffusion coefficients, ${\cal D}_{l_x x}^{l_x x}\neq{\cal D}_{l_y x}^{l_y x}(={\cal D}_{l_z x}^{l_z x})$ and, remarkably, enhances the overall orbital diffusivity. As explained in the main text, spin-orbit interaction couples the highly conductive spin channel with the weakly conductive orbital channel, thereby reducing the spin diffusivity while enhancing the orbital one, as shown in Fig. \ref{fig:cond}(c). These results are in qualitative agreement with the realistic calculations provided in the main text.

\begin{figure}[h!]
  \includegraphics[width=8.5cm]{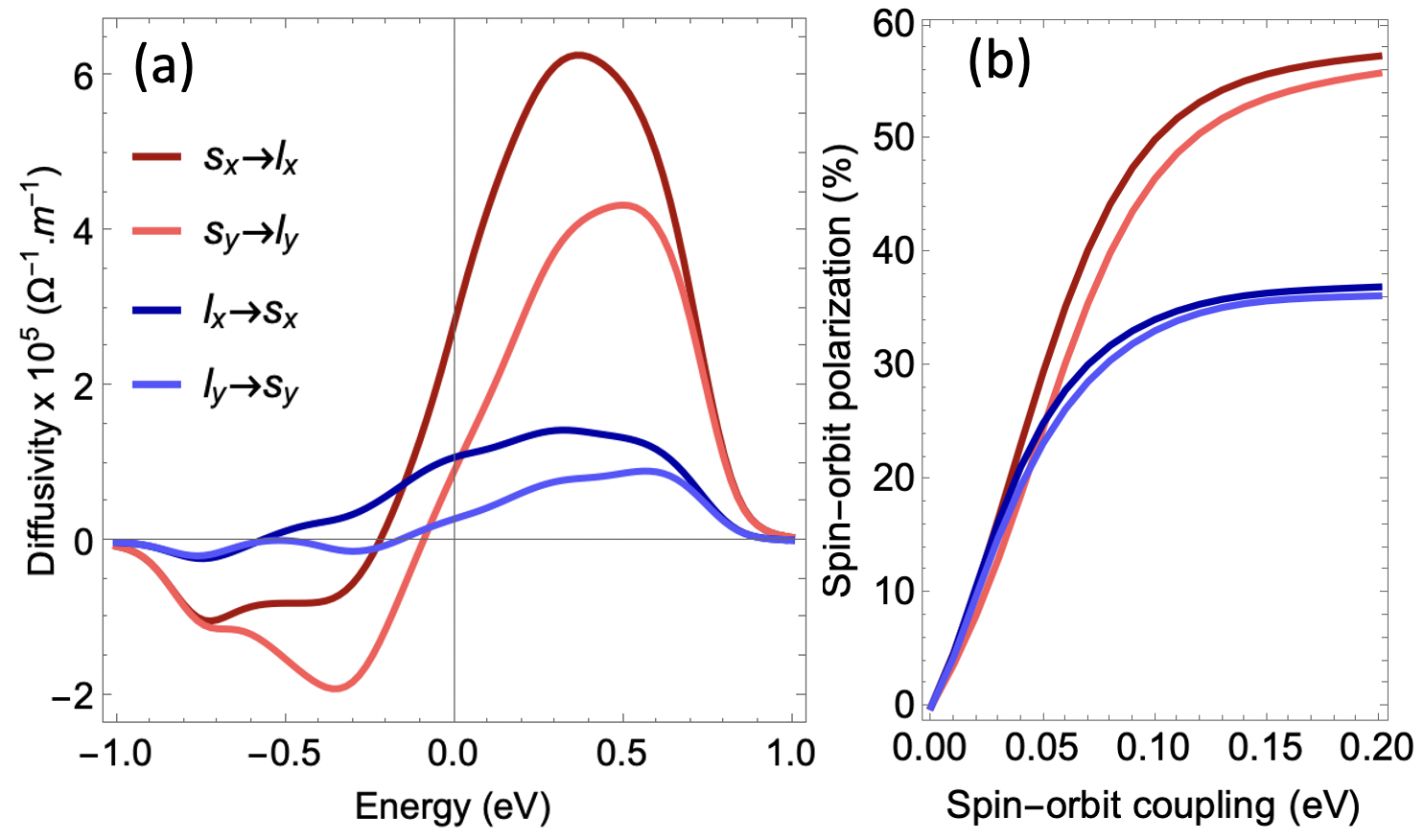}
  \caption{(Color online) (a) Spin-to-orbit (red) and orbit-to-spin (blue) diffusivities as a function of the energy for $\xi_{\rm so}=0.05$ eV. (b) Corresponding spin-orbit polarizations as a function of the spin-orbit coupling for $\varepsilon=0.5$ eV. \label{fig:socpol}}
\end{figure}

Let us now compute the "spin-orbit polarization" described in the main text. The spin-to-orbit ($s_\alpha\rightarrow l_\alpha$) and orbit-to-spin ($l_\alpha\rightarrow s_\alpha$) longitudinal diffusivities as well as the corresponding polarization are given in Fig. \ref{fig:socpol}(a,b). Again, the diffusivities are anisotropic due to the presence of spin-orbit coupling. The orbital diffusivity being much smaller than the spin diffusivity, the orbit-to-spin polarization is generally smaller than the spin-to-orbit polarization. The polarization increases steadily with spin-orbit coupling, as expected, and saturates at large spin-orbit coupling strength. The magnitude of the polarization is comparable to the one obtained in tantalum (see main text).

\begin{figure}[h!]
  \includegraphics[width=8.5cm]{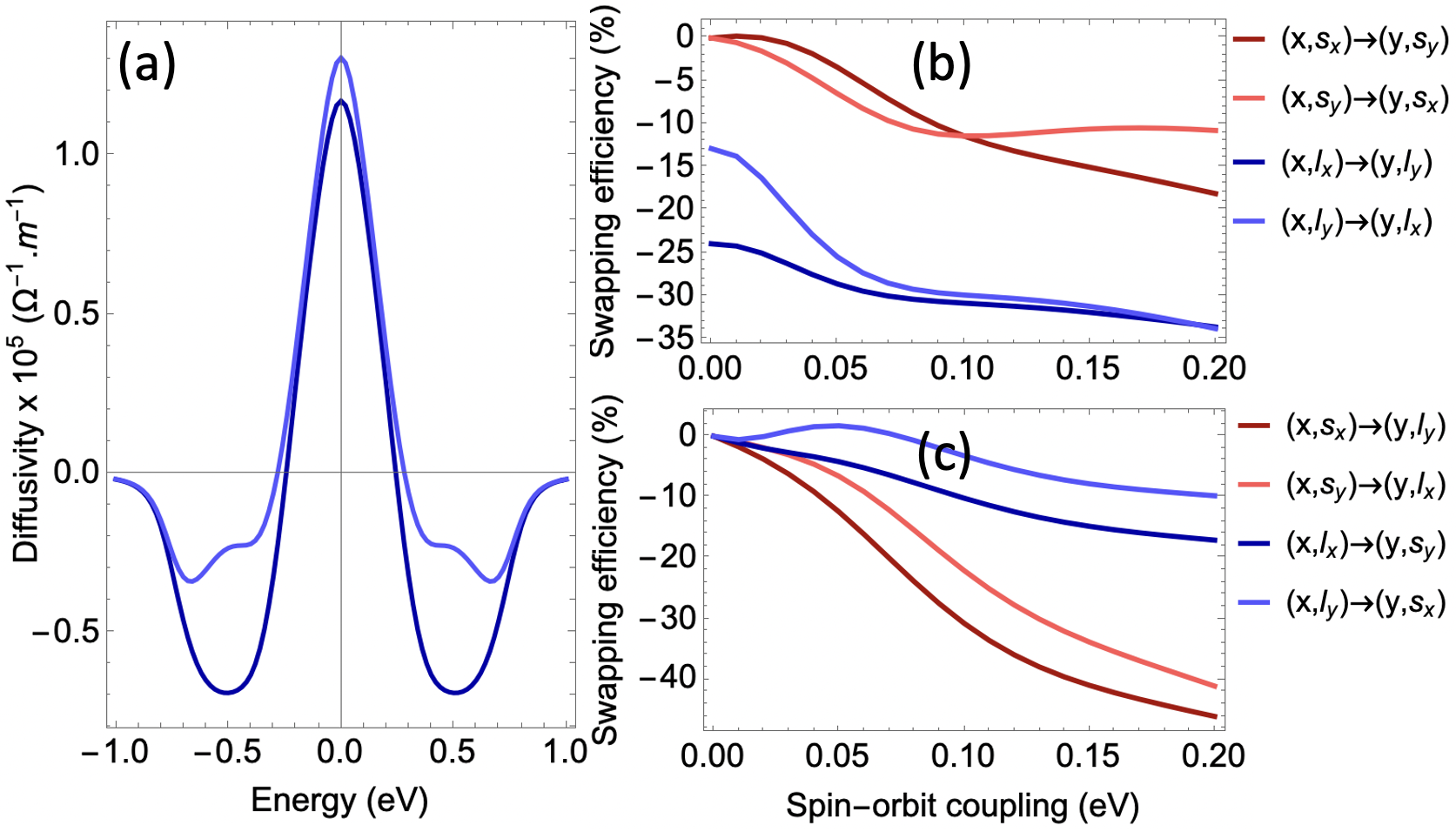}
  \caption{(Color online) (a) Orbital swapping as a function of energy for $\xi_{\rm so}=0$. (b) Spin (red) and orbital (blue) swapping as a function of the spin-orbit coupling strength. (c) Spin-to-orbit (red) and orbit-to-spin (blue) swapping as a function of the spin-orbit coupling. In (b) and (c), we set $\varepsilon=0.5$ eV. \label{fig:swapping}}
\end{figure}

We finally consider spin and orbital swapping. These coefficients are reported in Fig. \ref{fig:swapping}(a) in the absence of spin-orbit coupling. Figure \ref{fig:swapping}(b) displays the spin (red) and orbital (blue) swapping efficiencies defined as ${\cal D}_{s_xx}^{s_y y}/{\cal D}_{s_xx}^{s_x x}$ and ${\cal D}_{l_xx}^{l_y y}/{\cal D}_{l_xx}^{l_x x}$, as a function of spin-orbit coupling, showing that orbital swapping is generally larger than spin swapping, which seems reasonable given the minor role of spin-orbit coupling in the former. 

In addition, spin-orbit coupling also enables the transfer between spin and orbital angular momenta that results in spin-to-orbit (red) and orbit-to-spin (blue) swapping, displayed in Fig. \ref{fig:swapping}(c). From Fig. \ref{fig:swapping}(c), it appears that spin-to-orbit swapping is larger than orbit-to-spin swapping, a feature already observed in Fig. \ref{fig:socpol} for the spin-orbit polarization. 

\section{Two-channel model for orbital-spin interconversion}
Let us consider a phenomenological two-channel model, similar to the one famously used to explain spin transport in metals. Here, we consider the transport of angular momentum via the spin and the orbital channels. Each channel is characterized by its nonequilibrium distribution, $f_s$ and $f_o$, its relaxation time $\tau_s$ and  $\tau_o$, and a coupling rate $\tau_{so}$. In other words, the linearized Boltzmann transport equations reads
\begin{eqnarray}
({\bf v}\cdot{\bm\nabla})f_s=-\frac{1}{\tau_s}f_s+\frac{1}{\tau_{so}}(f_o-f_s),\\
({\bf v}\cdot{\bm\nabla})f_o=-\frac{1}{\tau_o}f_o-\frac{1}{\tau_{so}}(f_o-f_s).
\end{eqnarray}
Defining the spin and orbital densities as $n_{s,o}=\int d{\bf v}f_{s,o}$, and the spin and orbital current densities are ${\bf j}_{s,o}=\int d{\bf v}{\bf v}f_{s,o}$, we obtain the coupled equations
\begin{eqnarray}
\frac{v_s^2}{3}{\bm\nabla}n_s=-\left(\frac{1}{\tau_s}+\frac{1}{\tau_{so}}\right){\bf j}_s+\frac{1}{\tau_{so}}{\bf j}_o,\\
\frac{v_o^2}{3}{\bm\nabla}n_o=-\left(\frac{1}{\tau_o}+\frac{1}{\tau_{so}}\right){\bf j}_s+\frac{1}{\tau_{so}}{\bf j}_s.
\end{eqnarray}
After some manipulation, we end up with
\begin{eqnarray}
{\bf j}_s=-\frac{1+\frac{\tau_{so}}{\tau_o}}{1+\frac{\tau_s}{\tau_o}+\frac{\tau_{so}}{\tau_o}}D_s^0{\bm \nabla}n_s
-\frac{1}{1+\frac{\tau_o}{\tau_s}+\frac{\tau_{so}}{\tau_s}}D_o^0{\bm \nabla}n_o,\\
{\bf j}_o=-\frac{1}{1+\frac{\tau_s}{\tau_o}+\frac{\tau_{so}}{\tau_o}}D_s^0{\bm \nabla}n_s-\frac{1+\frac{\tau_{so}}{\tau_s}}{1+\frac{\tau_o}{\tau_s}+\frac{\tau_{so}}{\tau_s}}D_o^0{\bm \nabla}n_o.
\end{eqnarray}
We have defined the bare spin and orbital diffusivities as $D_{s,o}^0=v_{s,o}^2\tau_{s,o}/3$. From these definitions, we can deduce the spin, spin-to-orbital, orbital and orbital-to-spin diffusivities
\begin{eqnarray}
D_s&=&\frac{1+\frac{\tau_{so}}{\tau_o}}{1+\frac{\tau_s}{\tau_o}+\frac{\tau_{so}}{\tau_o}}D_s^0,\\
D_{so}&=&\frac{1}{1+\frac{\tau_s}{\tau_o}+\frac{\tau_{so}}{\tau_o}}D_s^0,\\
D_o&=&\frac{1+\frac{\tau_{so}}{\tau_s}}{1+\frac{\tau_o}{\tau_s}+\frac{\tau_{so}}{\tau_s}}D_o^0,\\
D_{os}&=&\frac{1}{1+\frac{\tau_o}{\tau_s}+\frac{\tau_{so}}{\tau_s}}D_o^0.
\end{eqnarray}
One immediately sees that the spin-to-orbital and orbital-to-spin conversion processes are inequivalent between orbital and spin moments as they exhibit different transport properties.

\section{Negative diffusivity}

The negative diffusivity of the orbital moment reported in Fig. \ref{fig:cond} and in the main text may seem rather unusual. It physically means that a gradient of orbital moment aligned on, say, ${\bf x}$, induces an orbital current polarized along $-{\bf x}$. The reason why such a change of sign occurs is that since the orbital moment does not commute with the Hamiltonian, it precesses around the crystal field and, therefore, can change sign upon diffusion. As pointed out by Ref. \cite{Han2022}, the orbital moment arises from a superposition of orbital-quenched states that propagate with different wave vectors. The dephasing between the orbital-quenched states results in the oscillation of the orbital moment during diffusion. 

\begin{figure*}[h!]
  \includegraphics[width=15cm]{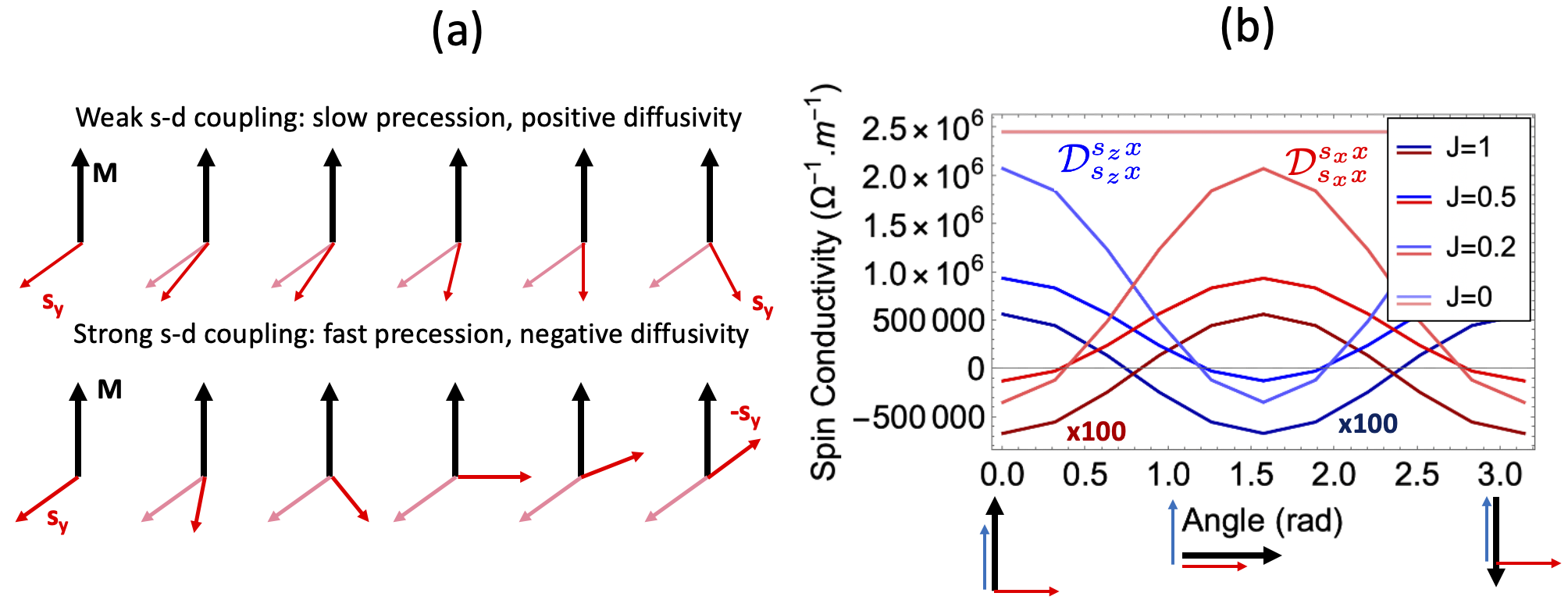}
  \caption{(Color online) (a) Schematics of the spin precessing occurring during the diffusion in the ferromagnet. (b) ${\cal D}_{s_xx}^{s_x x}$ (red) and ${\cal D}_{s_zx}^{s_z x}$ (blue) when rotating the magnetization, for various values of the s-d exchange energy. The case $J=1$ has been multiplied by 100. \label{fig:Fig5}}
\end{figure*}

A similar effect is observed for the electron spin in a ferromagnet. Due to the s-d exchange, the polarization of a spin current precesses around the magnetization as soon as it is misaligned with it \cite{Zhang2002,Petitjean2012,Pauyac2018}. To illustrate this effect, we used the model presented in section IV of this Supplemental Material and added a magnetic exchange, $J{\bm\sigma}\cdot{\bf m}$, keeping all the other parameters unchanged. We computed the spin diffusivities ${\cal D}_{s_xx}^{s_x x}$ and ${\cal D}_{s_zx}^{s_z x}$ when rotating the magnetization in the ($x,z$) plane, from $+{\bf z}$ to $-{\bf z}$. The mechanism is depicted in Fig. \ref{fig:Fig5}(a) and the results are shown in Fig. \ref{fig:Fig5}(b).

When the magnetization points toward $+{\bf z}$, ${\cal D}_{s_zx}^{s_z x}$ is always positive, whereas ${\cal D}_{s_xx}^{s_x x}$ changes sign when increasing the exchange. As illustrated in Fig. \ref{fig:Fig5}(a), for a weak exchange, the precession is slow and the incoming spin only slightly precesses over the diffusion length. As a result, the diffusion coefficient is positive. For a strong exchange, the precession is fast and the incoming spin precesses strongly over the diffusion length, ending up being completely reoriented. As a result, the diffusion coefficient is negative. Notice that increasing the exchange also reduces the magnitude of the diffusivities. In contrast, when the magnetization points toward $+{\bf x}$,  ${\cal D}_{s_xx}^{s_x x}$ remains positive for all values of the exchange, whereas ${\cal D}_{s_zx}^{s_z x}$ changes sign. Finally, we note that switching to $-{\bf z}$ gives the same result as $+{\bf z}$, which is reasonable since the handedness of the precession does not matter.

\bibliography{Biblio2023}